\documentclass[conference,10pt,compsocconf]{IEEEtran}
\usepackage{setspace}
\usepackage[utf8]{inputenc}
\usepackage{tikz}
\usepackage{amsmath}
\usepackage{graphics}
\let\labelindent\relax
\usepackage{enumitem}
\usepackage{newfloat}
\usepackage{float}
\usepackage[none]{hyphenat}
\usepackage{dblfloatfix}
\makeatletter
\renewcommand\footnoterule{%
  \kern-3\p@
  \hrule\@width 0.5\columnwidth
  \kern2.6\p@}
  \makeatother
\begin{document}
\title{\Large{\textbf{Side-Channel Attacks on RISC-V Processors:\\ Current Progress, Challenges, and Opportunities}}\vspace{-20pt}}
\author{
 \IEEEauthorblockN{Mahya Morid Ahmadi$^{1}$,
        Faiq Khalid$^{1}$, 
 		Muhammad Shafique$^{2}$}
  	\IEEEauthorblockA{$^1$\textit{Technische Universit\"at Wien (TU Wien), Vienna, Austria}\\
  	$^2$\textit{Division of Engineering, New York University Abu Dhabi (NYUAD), Abu Dhabi, United Arab Emirates}\\
  	Email: \{mahya.ahmadi,faiq.khalid\}@tuwien.ac.at, muhammad.shafique@nyu.edu}\\ \vspace{-20pt}
 }
\maketitle
\begin{abstract}
Side-channel attacks on microprocessors, like the RISC-V, exhibit security vulnerabilities that lead to several design challenges. Hence, it is imperative to study and analyze these security vulnerabilities comprehensively. In this paper, we present a brief yet comprehensive study of the security vulnerabilities in modern microprocessors with respect to side-channel attacks and their respective mitigation techniques. The focus of this paper is to analyze the hardware-exploitable side-channel attack using power consumption and software-exploitable side-channel attacks to manipulate cache. Towards this, we perform an in-depth analysis of the applicability and practical implications of cache attacks on RISC-V microprocessors and their associated challenges. Finally, based on the comparative study and our analysis, we highlight some key research directions to develop robust RISC-V microprocessors that are resilient to side-channel attacks.
\end{abstract}
\section{Introduction}
\label{Introduction}

\begin{figure}[!t]
\centering
\includegraphics[width=1.0\linewidth]{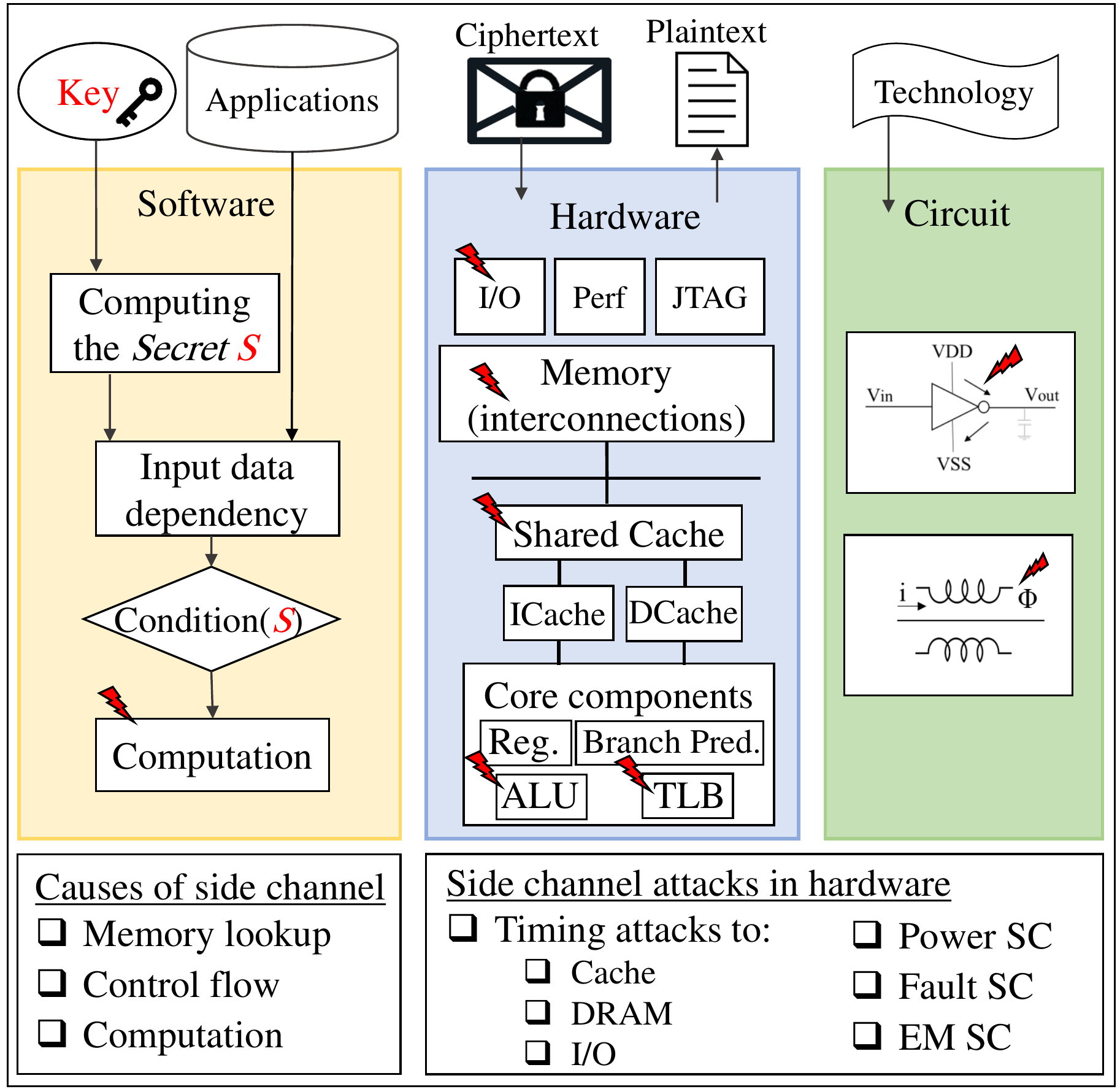}
\vspace{-10pt}
\caption{Security vulnerabilities in microprocessors that can be exploited by side-channel data leakage attacks, i.e., timing attack to cache, memory and I/O, and power, fault-based, and electro-magnetic SC attacks. }
\vspace{-5mm}
\label{fig:Src}
\end{figure}

The exponential increase in using advanced microprocessors for critical applications (e.g., surveillance systems) makes these systems vulnerable to several security threats, e.g., remote micro-architectural~\cite{Survey1} and side-channel attacks~\cite{Ge}. These attacks may lead to system failure, information leakage, and denial-of-service. Therefore, it is imperative to study security as a fundamental parameter along with performance constraints in the early design stages of these microprocessors, especially the emerging microprocessors like in fifth generation of Reduced Instruction Set Computer (RISC-V). To address this critical issue in microprocessors, researchers have developed several defenses against these threats. However, broadly, research in the security of microprocessors has been mostly divided into independent directions of hardware and software threats~\cite{CPS1}\cite{CPS2}. In hardware security, researchers are focused on chip modification and physical intrusions~\cite{Survey2}, and at the software level, they are studying software stack attacks like software malware~\cite{ye}. Therefore, these are not applicable to the advanced attacks, such as the software-exploitable hardware attacks. These attacks combine the software and hardware bugs to exploit the microprocessor at run-time. For example, software-exploitable timing side-channels~\cite{Ge}\cite{Lyu} has been exposed only a couple of years ago. As shown in Figure~\ref{fig:Src}, the execution of software leaves side-channel traces at different levels of the system. These traces can be exploited by software-exploitable side-channel attacks to remotely leak confidential information from the trusted hardware with high bandwidth microarchitectural channels. Typically, these attacks target shared resources like last-level cache, which are inevitable in the high performance emerging microprocessors~\cite{Survey3}. Recent studies show that these vulnerabilities are not limited to current microprocessors, but the next generation of microprocessors are also vulnerable to these attacks. Hence, it is imperative to study these vulnerabilities in emerging microprocessors, like RISC-V, to make them robust against these powerful attacks.

\begin{figure*}[!b]
\centering
\includegraphics[width=0.95\textwidth]{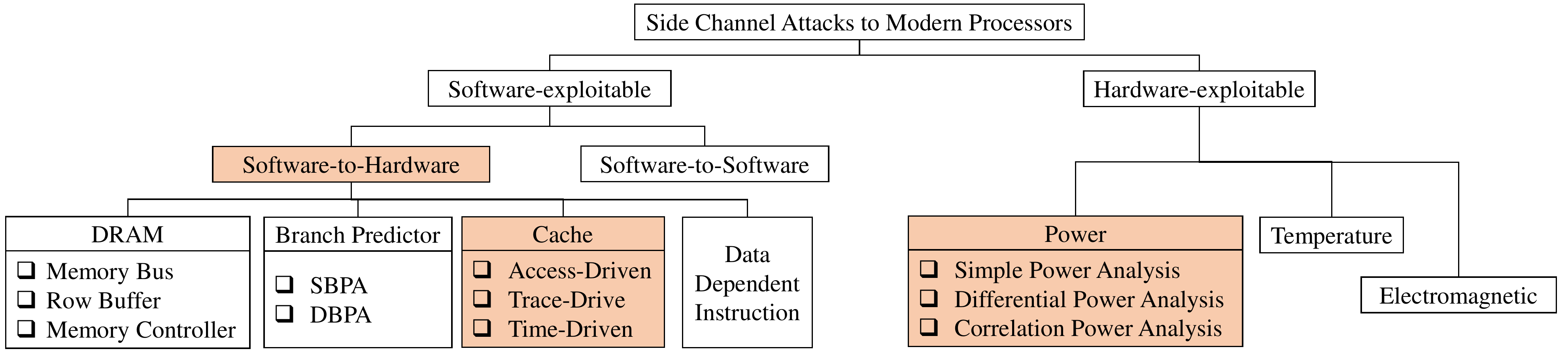}
\vspace{-10pt}
\caption{Taxonomy of side-channel attacks to processors. The main focus of this paper is to study software-to-hardware side-channel attacks, that exploited as timing side-channels in cache, and power side-channel attacks, as shown by the highlighted box. In this figure, SBPA and DBPA represent the simple branch prediction analysis and dynamic branch prediction analysis, respectively.}\vspace{-15pt}
\label{fig:SCA-DEF} 
\end{figure*}

Therefore, the main focus of this paper is to highlight the security vulnerabilities in one of the most important emerging microprocessors, i.e., RISC-V. The reason behind this is that RISC-V is predicted to be a widely-adopted architecture in the coming days, as its rapid proliferation has already been witnessed in both industrial and academic Research and Development (R\&D) projects and product lines~\cite{risc-v-imp}. Moreover, RISC-V~\cite{risc-v}, as an open-source, with extensible Instruction Set Architecture (ISA), is rapidly becoming the mainstream architecture for emerging embedded systems. The flexibility of this ISA made it popular in lightweight embedded systems of edge devices as well as server-side complex multi-core systems. Open hardware designs are essential for security, low-power applications, and fast R\&D cycles. Since the RISC-V microprocessors are in the early stage of R\&D, therefore, it is the right time to investigate security solutions there, with an eye to the past mistakes to avoid them in the next-generation of processors. To protect emerging embedded systems against security vulnerabilities, first, the applicability of the state-of-the-art software-exploitable attacks on RISC-V processors must be studied.  Then, based on the observations of these studies, a new attack surface can be found that can exploit the unique features of RISC-V ISA (e.g., memory model). Although researchers have developed several security solutions for RISC-V, their primary focus is on the software attacks. However, there are a few security solutions for software-exploitable hardware attacks on RISC-V, which leads to a key research question about \textit{how to design a robust RISC-V microprocessor that can tolerate software, hardware, and software-exploitable hardware attacks?}

Towards the above-mentioned research question, in this paper, \textbf{we made the following key contributions}: 
\begin{enumerate}[leftmargin=*]
    \item We also provide a brief yet comprehensive overview and categorization (\textbf{Section~\ref{sec:side-channels}}) of the different side-channel attacks for microprocessors and their respective defenses(\textbf{Section~\ref{sec:defens}}). 
    \item We first study the side-channel data leakage attack on RISC-V processors running confidential applications like RSA encryption (\textbf{Sections~\ref{sec:SC-power} and~\ref{sec:SC-timing}}). 
    \item Based on the study, we demonstrated the architectural vulnerabilities in a high-performance RISC-V microprocessor by successfully implementing a cross-core timing side-channel attack on the RISC-V microprocessors (\textbf{Section~\ref{sec:SC-timing_riscv}}). Our results show that RISC-V is vulnerable to timing side-channel attacks; hence, a lightweight yet powerful defense mechanism is required. 
    \item Towards the end, we briefly discuss the potential defenses for the cross-layer side-channel attacks on the RISC-V microprocessor (\textbf{Section~\ref{sec:riscv-def}}) and highlighted research challenges and opportunities for cross-layer side-channel attacks on a RISC-V microprocessor (\textbf{Section~\ref{Challenges}}). 
\end{enumerate}

\section{Background}
\label{Background}
In this section, we provide the necessary background information for side-channel attacks on RISC-V microprocessors. First, we presents the taxonomy for state-of-the-art side-channel attacks with the focus on power side-channel attacks and timing attacks. Then, we discuss the vulnerabilities in the encryption algorithms, and towards the end, we present a brief overview of the RISC-V ISA.

\subsection{Side-Channel Attacks}\label{sec:side-channels}
There is a large body of the work on hardware attack to microprocessors~\cite{Ge}. However, in this paper, we focus on attacks in hardware that are exploiting an unwanted channel to leak confidential data, which are known as side-channels. The principle of a security manipulation starts with forcing the victim to transfer the critical data to the target location, then store it or consume it, which gives the opportunity to attacker to steal the data. If the leaking channel of data is built using physical parameters of the system, which is dependant on the control flow, it is called a side-channel. There are several side-channels reported in hardware, and among the most exploited ones, we can name power, timing, and temperature.

In Figure~\ref{fig:SCA-DEF}, we are presenting a known taxonomy of side-channel attack and their mitigation techniques in general-purpose microprocessors. Based on the attack model, attacks are divided into two categories: sourced from software and sourced from the hardware. Attacks from hardware exploit physical features, e.g., dynamic power, by implanting sensors in shared resources of the system. While the former category requires physical access to the microprocessor, attacks from software can be exploited remotely. Side-channel attacks from software leak the confidential data through microarchitectural events of shared resources, e.g., timing attack to cache~\cite{cache}. 
 As the efforts on the side-channel attack to RISC-V processors are focused on power and timing attacks, we are elaborating on these attacks in the following sections.
 
\begin{figure}[!t]
\centering
\includegraphics[width=1.0\linewidth]{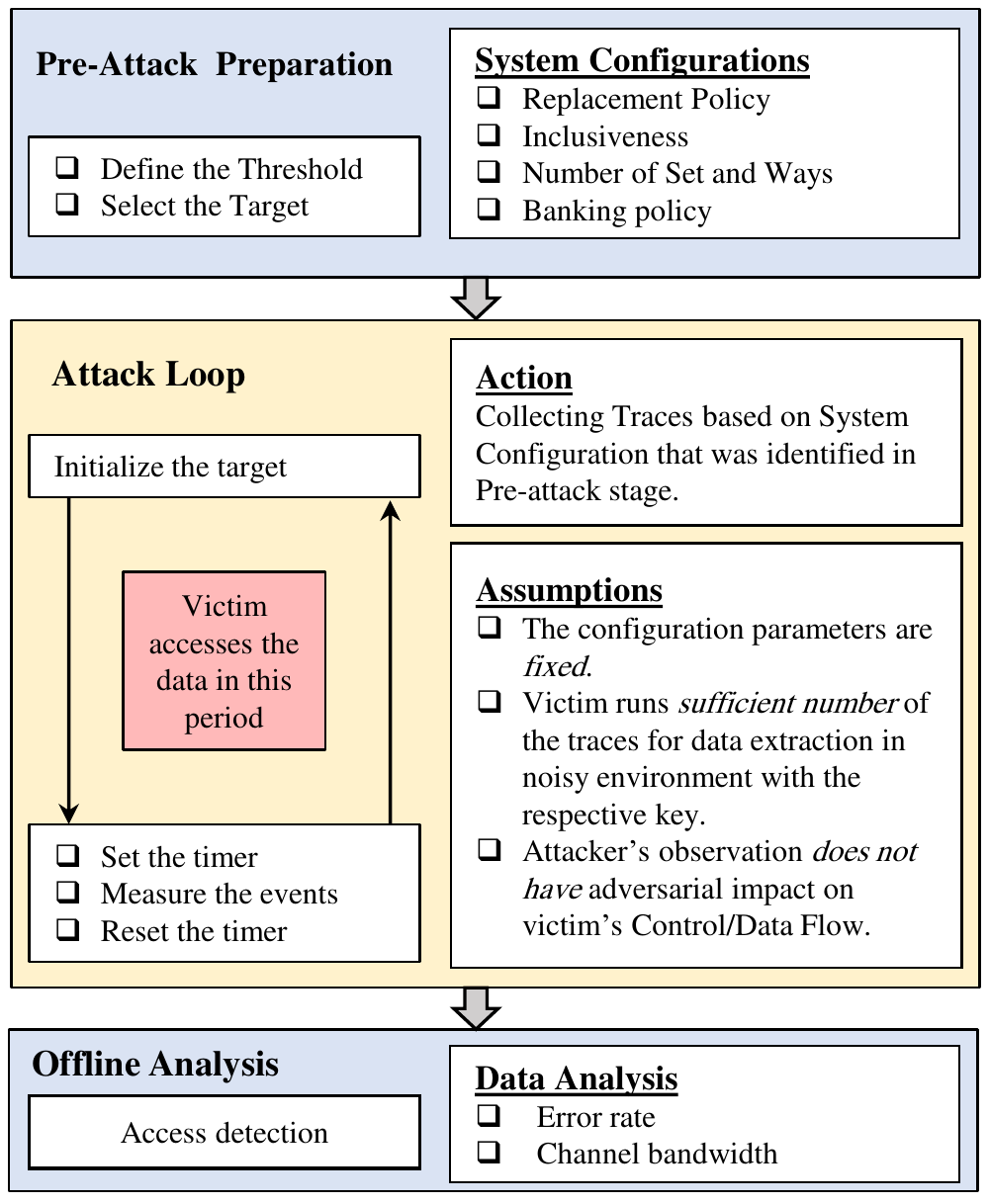}
\vspace{-10pt}
\caption{Principles of timing attacks to shared components. Each timing attack, is consist of three stages: \textbf{ Pre-attack}, \textbf{attack loop} and \textbf{data analysis}.}
\label{fig:TimingAtt}
\vspace{-5mm}
\end{figure}

\begin{figure*}[!t]
\centering
\includegraphics[width=0.95\textwidth]{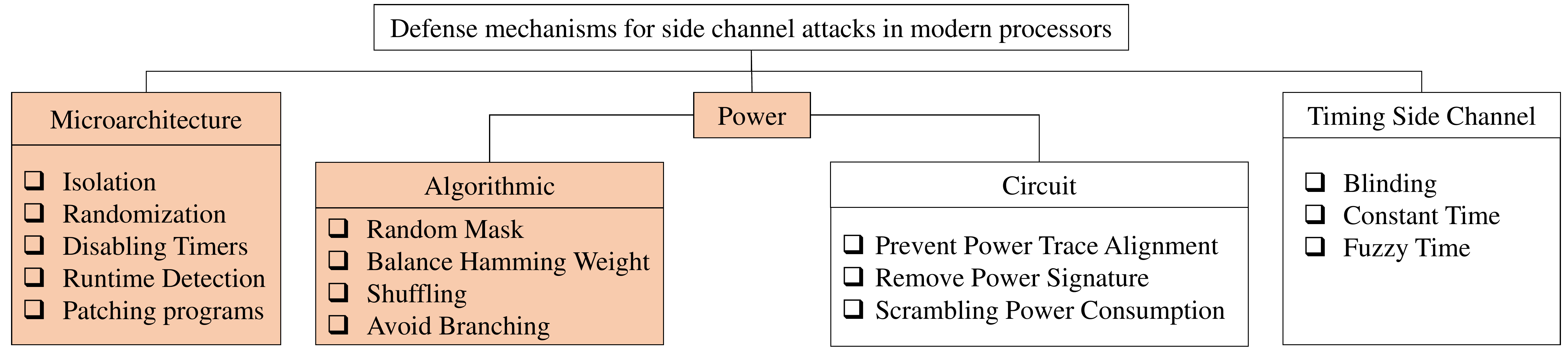}
\vspace{-10pt}
\caption{A taxonomy of defense techniques against side-channel attacks in the processors is presented. Based on the defense point of application, mechanisms are divided into 3 categories. In this paper, we give an overview on defense mechanisms in power and microarchitecture categories.}\vspace{-5pt}
\label{fig:TAX-DEF} 
\end{figure*}

        \subsubsection{\textbf{Power attacks}}
       In cryptographic algorithms, the computation is based on a secret value \textit{S}, which is not necessarily the key to encryption but it is derived from the key. The activity of encryption software causes data dependant signal transitions in the current surges of the hardware, which are visible in internal wires, power, and ground connection. By collecting the dynamic power traces, an adversary can find the secret value correlated to the input data patter while considering measurement errors and enviornmental noise. 
       \par 
       Power traces built on the direct current correspondence of the software operation, are analyzed as Simple Power Analysis (SPA). In this analysis it is assumed that a change in each bit of \textit{S} is visible in the leakage power pattern. In attack to applications with complicated relation of input and application flow, Differential Power Analysis (DPA) can reveal the \textit{S} by statistical tests on the differentiation of many power traces. DPA is a very powerful technique that can easily eliminate the random noise sourced from environment or obfuscation mitigation techniques.

        \subsubsection{\textbf{Timing attacks}}
         Microarchitectural side-channels are typically timing-based channels, as shown in Figure~\ref{fig:TimingAtt}. Due to sharing functional units among different programs, an attacker can, in general, observe the timing of the operation of the individual functional unit and its output. Since the designs of the functional units are knows to the attacker, these timing leakages reveal whether the fast or slow execution path was taken~\cite{Szefer}. 
         In the taxonomy presented in Figure~\ref{fig:SCA-DEF}, Software-to-hardware side-channel attacks are categorized based on the shared unit. In~\cite{Predictor}, the attacker manipulates a shared branch predictor to infers victim's execution path.
         Cache, as an inevitable unit of microprocessor, has been the target of side-channel attacks. Cache timing attacks are based on the contention of lines between victim and attacker or data-reuse in the application. They can exploit strong isolation techniques from the core level to threads. 
    
    With Osvic~\cite{Osvic} proposing several exploitation techniques, the timing attacks can be categorized as Time-Driven and Access-Driven, based on the source of leakage, an attacker's access, and application. Later on, trace-driven attacks were also recognized as a separate category. 
    
    \begin{itemize}[leftmargin=*]
     \item \textit{Access-Driven Attacks:} These attacks are independent of the victim's performance counters, but they measure the effect of spatial and temporal resource sharing between the attacker's application and victim's application. The pieces of evidence in the attacker's workspace about the victim's access are used to reveal the encryption key. Since this attack leaves no trace of access, it is considered more stealthy. {\it PRIME+PROBE}~\cite{Prime}\cite{b2-0} is a cross-core and cross-VM Access-Driven attack, which finds the pattern of victim application memory access without the need to flush the cache or assuming shared addresses in the memory.
    {\it FLUSH+RELOAD}~\cite{Flush} is another example where the attacker shares an address in the memory with the victim. The attacker benefits the shared virtual memory such as page deduplication or shared libraries managed by the operating system to make contention with the victim. In this attack, the attacker tries to flush the shared addresses and measure his second access to find the victim's internal execution path. Although this attack has been exploited using several variants, the shared address assumption limits the threat model. Also, there is a need for an atomic instruction for flushing a specific line which is not present in all architectures (e.g., RISC-V). As another example is {\it FLUSH+FLUSH}~\cite{Flush2}, where malicious user records the loop time for {\it clflush} instruction in the attack-loop phase.
    
    \item \textit{Time-Driven Attacks:} In this category of attacks, attacker attempts to measure the victim's execution time. The execution time of applications is dependant on several parameters like execution path, data flow, and memory access time. 
    Hence, these attacks require a large and detailed execution profile for key extraction. {\it EVICT+RELOAD}~\cite{Evict} is a Time-Driven cache attack, whereby evicting specific cache lines, the attacker prepares the system to reveal the victim's control flow. By forcing the victim to run enough times, attacker gains the required number of traces to extract key-dependent memory accesses. This attack is based on the average execution time of a confidential program. Therefore, it needs a large number of sample data to recognize the key.
    
    \item \textit{Trace-Driven Attacks:} In this category, the attacker collects the traces of encryption on a shared resource, e.g., cache, based on a known message attack. These traces can form a profile of hit and miss in the cache, which gives information about the key-dependent lookup addresses. This ability gives an adversary the opportunity to make inferences about the secret key. In \cite{Trace}, cache traces are determined by the power consumption in the cache for each hit and miss to break the AES encryption algorithm. 
    \end{itemize}

\subsection{Defense Mechanisms}\label{sec:defens}
In Figure~\ref{fig:TAX-DEF}, defense mechanisms are categorized into three subsets, based on the source of leakage channel they are protecting against. Microarchitectural mitigation techniques, e.g., isolation, are implemented both in software and hardware levels that each can address a specific range of attacks \cite{Scache}. Also, mitigation techniques against power side-channels, are proposed both at circuit level, as resilient processors, and algorithmic level, as resilient application implementation. We will present works in these categories in Section IV.

\subsection{Data Dependant Implementation of Encryption Algorithms}
\begin{figure}[h]
\centering
\includegraphics[width=0.55\linewidth]{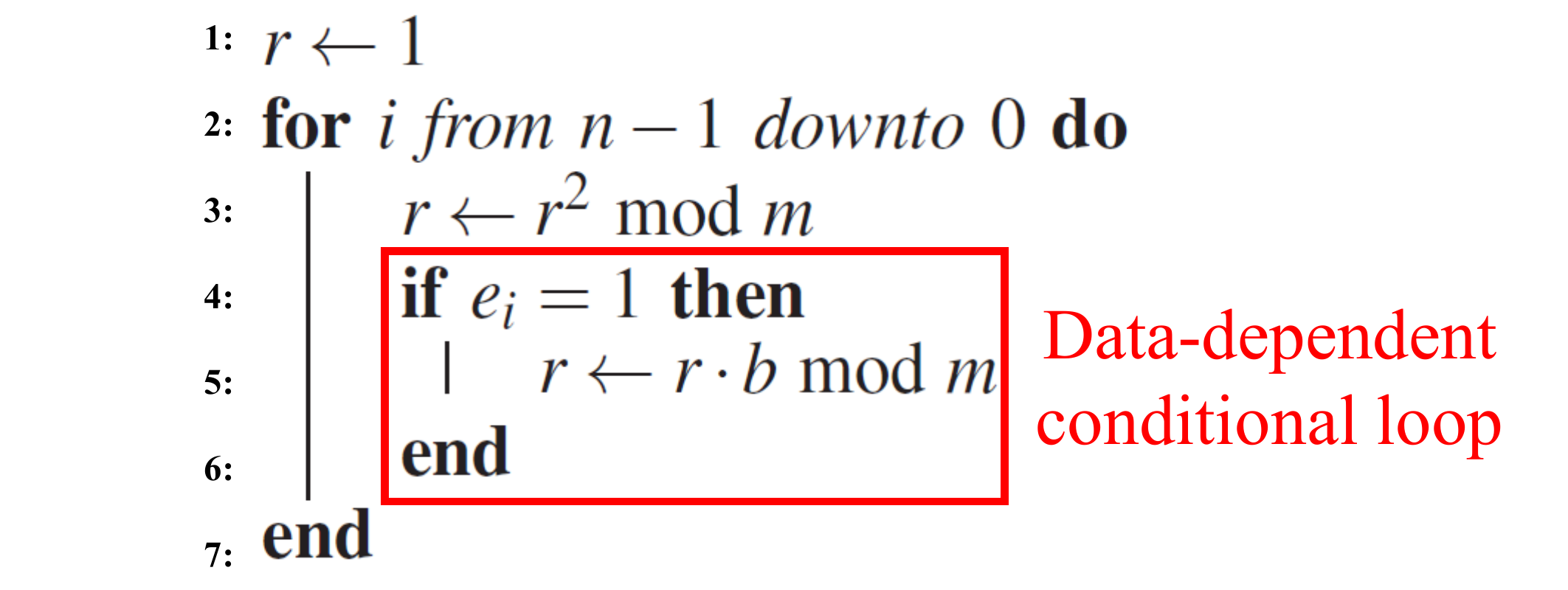}
\caption{Data-dependent implementation of modular exponentiation which is used in RSA encryption algorithm.}\vspace{-5pt}
\label{fig:Encryption}
\vspace{-2mm}
\end{figure}

In an attack on cryptographic algorithms, the encryption key is the target of the attacker. Two implementation models are mostly analyzed in the literature for the security of encryption application implementation, conditional control flow, and conditional lookup table, which are respectively used in Rivest–Shamir–Adleman (RSA) and Advanced Encryption Standard (AES) encryption systems. 
\begin{enumerate}[leftmargin=*]
    \item \textit{Conditional Control Flow:}
    RSA is a cryptosystem for exchanging the key between two parties. To calculate the private key and the public key, modular exponentiation is implemented using a square-and-multiply algorithm that computes $r = b^e~mod~m$ dependant to bits of $e$ in a conditional loop. As shown in Figure \ref{fig:Encryption}, line 5 will be executed if the condition in line 4 is fulfilled and it causes bit-dependant physical footprints like dynamic power consumption and execution time. In~\cite{RSA}, Mushtaq et al. have presented a comprehensive survey in attacks and countermeasures on RSA.
    
    \item \textit{Conditional Table Lookup:}
    In block cipher cryptosystems, the secret of the system is not directly used for encryption, but the key enables an address in the pre-computed tables of data. When the index used to access a Look-Up Table (LUT), depends on the secret data, the access time and dynamic power consumption may vary due to the behavior of memory access. Such data leakages have been exploited in various block ciphers, e.g., AES in \cite{AES} that implements S-Boxes using lookup tables to reduce hardware overhead.
\end{enumerate}

\subsection{RISC-V Architecture} 
\begin{figure}[!t]
\centering
\includegraphics[width=1.0\linewidth]{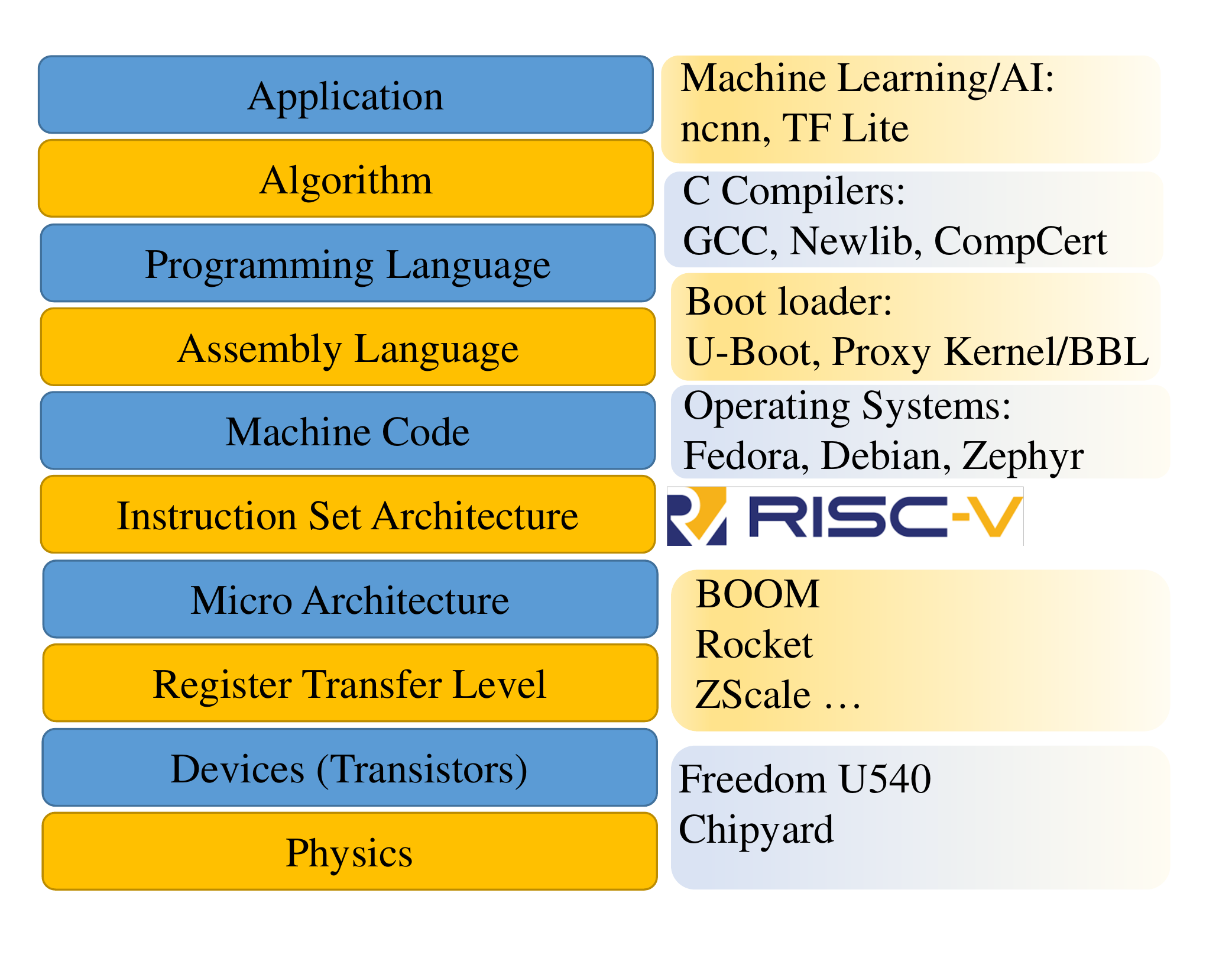}
\vspace{-30pt}
\caption{RISC-V ecosystem is shown. RISC-V implements the ISA and connects software stack to hardware. In each layer some examples, of the current available implementations are noted. It should be mentioned that there are other products in the market and academia which are not named here.}
\label{fig:Ecosystem}
\end{figure}

RISC-V~\cite{risc-v} with open-source ISA is received as an alternative in academia and also popular in the IoT industry. Since it is a flexible, upgradeable, and optimizable architecture, it is predicted to be adopted widely in emerging embedded systems. 
RISC-V ISA is not limited to a specific hardware implementation of a processor core. The modular architecture implementation with variants of address space sizes makes it suitable for lightweight edge devices to the high-performance servers. The base integer instruction set is flexible to adopt the required software stack and to add additional requirements. Moreover, the simplicity of the design enables the RISC-V to be an ideal base for the application-specific design. Accelerators and co-processors can share the compiler tool-chain, operating system binaries, and control processor implementations. It speeds up the process of agile hardware processor design by the support of an active and diverse community for open-source contributions. Figure~\ref{fig:Ecosystem} shows an overview of the RISC-V ecosystem and respective tools.


\section{Side-channel Attacks on RISC-V Microprocessors}\label{sec:SC-RISCV}
In this section, we discuss state-of-the side-channel attacks on RISC-V microprocessors. 
\begin{figure*}[!b]
\centering
\includegraphics[width=1.0\textwidth]{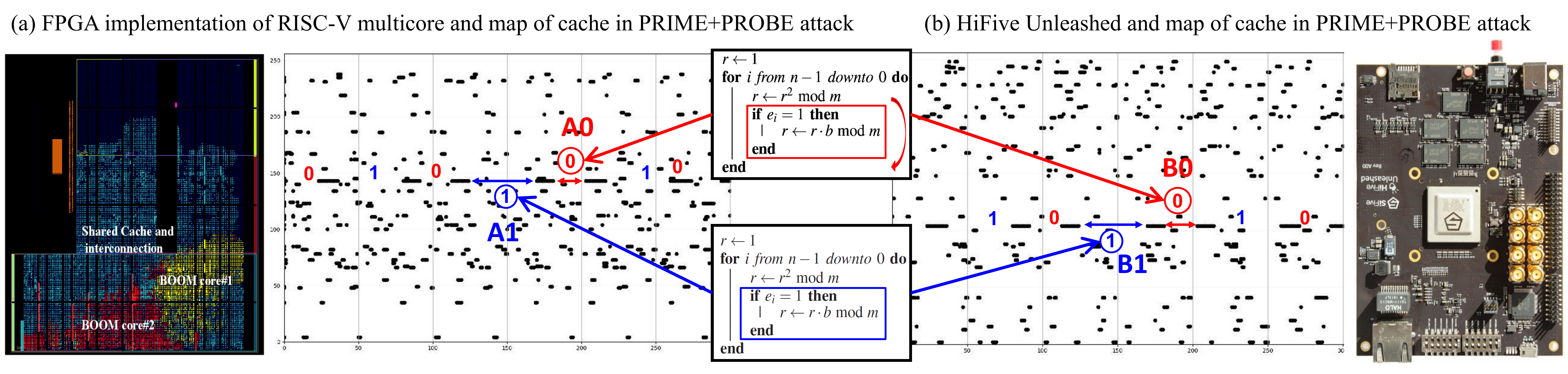}
\vspace{-20pt}
\caption{ {\it PRIME+PROBE} cache timing attack on RISC-V hardware platforms which presents the content map of the cache for \textbf{300} time slots, when our attack is priming \textbf{256} sets. The access pattern reveals the keys of RSA algorithm. The long empty interval means victim's data was processing and it was present in the cache, while short empty interval indicates that the loop was not executed and victim releases the target set. Note, in this figure, Where the key bits are 1, the cache intervals are longer than key bits that are 0 because of the data dependant conditional flow in the RSA Algorithm.}
\label{fig:Results}
\vspace{-1mm}
\end{figure*}

\subsection{Power Side-Channel Attacks}\label{sec:SC-power}
    Mulder et al. in ~\cite{Mulder} discuss power side-channel vulnerabilities of RISC-V microprocessor. In this work, they distinguish between different types of data leakage by categorizing them into direct-value leaks, data-overwrites and circuit-level leaks. Since direct-value leakage can be protected using masking techniques, they show software masking is not sufficient for mitigating power side-channel. On the other hand, since side-channel analysis is usually postponed to post-fabrication stages of hardware production, the state-of-the-art is focused on studying the software implementation of cryptographic algorithm. Therefore, a promising research direction is to apply the mitigation technique in the architectural level to prevent exceeding complexity of the microprocessor-based systems. In this work, they exploit memory access patterns of an AES encryption algorithm through a data overwrite leakage. This leakage happens when the data and the mask are accessed repetitively through memory connections. The replaced data in intermediate registers causes a peak in consumed power, which is leaked by power side-channel attack.  
    \subsection{ Timing Side-Channel Attacks}\label{sec:SC-timing}
    Recent attacks on modern processors have shown that special features of emerging architectures can be a source of the attack in lower hardware levels, as well as in the previous generation. In \cite{Gonzalez}, Gonzalez et al. replicated Spectre~\cite{Spectre} attack on BOOM (Berkeley out-of-order-machine) core and exploits in-core data cache for leaking the confidential data. Since the \textit{clflush} instruction is not implemented in the RISC-V ISA, the authors implement an atomic instruction to replace the targeted lines with dummy data and evict the victim's shared page. More recently, Le et al. in \cite{Carrv}, showed a side-channel attack to physically-implemented BOOM. In this attack, they target the conditional branch, which is trained to execute an instruction that assesses the confidential data.
    
    \section{{\it PRIME+PROBE} attack on RISC-V}\label{sec:SC-timing_riscv}
    To identify the vulnerabilities of RISC-V against timing side-channel attacks, we implemented a state-of-the-art cache timing attack ({\it PRIME+PROBE}) to RSA encryption algorithm running on different RISC-V hardware platforms, i.e., an out-of-order speculative RISC-V core (BOOM, implemented as SoC on the FPGA) and a commercial in-order RISC-V CPU (HiFive Unleashed).

    \subsection{Design Challenges}
    Most of the timing side-channel attacks are applicable to traditional microprocessors' architecture. However, the implementation of these attacks on the RISC-V microprocessors exhibits the following design challenges:
\begin{enumerate}[leftmargin=*]
    \item The cache addressing is not determined by RISC-V ISA, and it differs with implementation. 
    For instance, the last-level cache addressing for Rocket SoC is designed to be virtually-indexed, which is the one that helps to build a shared set between the victim and the attacker. 
    \item In this work, we replicate the attack on a high-performance commercial processor to find out the impact of branch prediction and out-of-order units on the timing of application execution. For instance, if the attacker accesses to one set in the cache, the prefetch unit can learn the access pattern and predict the execution in pause times, while shuffling the lines in this attack prevents this.
    
    \item The RISC-V ISA does not support the \textit{clflush} instruction that is exploited by cache timing attacks in Intel microprocessors (\textit{FLUSH+RELOAD}). Therefore, we choose another attack that is independent of special instruction or shared addresses in the victim's software space. We observe that RISC-V microprocessors reveal secrets of the system with the same trend in bandwidth as commercial microprocessors.
\end{enumerate}

\subsection{Experimental analysis on hardware platforms}
We performed the {\it PRIME+PROBE} attack on a Rocket SoC with BOOM RISC-V multi-core implemented on the Zedboard, Xilinx Zynq-7000 FPGA board, and we show the successful implementation of the cache timing attack ({\it PRIME+PROBE}) to the RSA encryption Algorithm on RISC-V ISA in Figure~\ref{fig:Results}(a). For the comprehensive analysis, we also performed the {\it PRIME+PROBE} attack on HiFive Unleashed, i.e., a commercial RISC-V multi-core CPU (see Figure~\ref{fig:Results}(b)). Since, in the {\it PRIME+PROBE} attack, the attacker makes consecutive accesses to targeted cache lines in each determined time slot. By measuring the access time, the attacker can decide whether the content of the requested address is present in the cache or it has been replaced by the victim's access. By analyzing the results in Figures~\ref{fig:Results}(a) and (b), we made the following key observations:  

\begin{enumerate}[leftmargin=*]
    \item The difference between cache access time for key-bit 0 (see labels A0 and B0 in Figure~\ref{fig:Results}) and key-bit 1 (see labels A1 and B1 in Figure~\ref{fig:Results}) is distinguishable even when other processes are running on the shared platform. In our attack implementation, we have exploited this timing differences to extract the key.
    \item The identified timing vulnerabilities exist in the both FPGA implementation and HiFive Unleashed board that can be exploited by the {\it PRIME+PROBE} attack. Hence, these vulnerabilities are independent of the hardware platform. This observation leads us to a conclusion that mitigation technique for FPGA implementations can also be deployed ASIC implementation of RISV-V microprocessors.
    \item These timing vulnerabilities can be exploited to perform other software-exploitable attacks on the RISC-V hardware, for example, Spectre~\cite{Spectre} and Meltdown~\cite{Meltdown}.
\end{enumerate}
\section{Defense mechanisms for Side-channel Attacks on RISC-V Microprocessors}\label{sec:riscv-def}
The usage of high-performance processors, e.g., commercial RISC-V CPUs~\cite{Gonzalez}, in Internet-of-Things (IoT) devices, have already been shown to be vulnerable to microarchitectural attacks like Spectre~\cite{Spectre} and Meltdown~\cite{Meltdown}. Efforts in defense techniques are focused on protecting sensitive data in memory locations ~\cite{Sanctorum}\cite{Shakti}, and protecting the software by providing an isolated execution environment~\cite{TIMBER-V}\cite{Keystone}. Yu et al in~\cite{OISA} discuss a compiler-based solution for securing data oblivious code generation for preventing the side-channel attack, which is not applicable at software level. To protect the RISC-V CPU against power side-channel attacks, Mulder et al.~\cite{Mulder} have proposed a masking solution at the architecture level.

\section{Open Research Challenges and road ahead}
\label{Challenges}
    
Currently, the focus of research in the security of RISC-V is on the security of the software. 
While it is crucial to protect software integrity, hardware security is a significant threat that needs to be addressed accordingly. In the following, we mention two main challenges in this research direction.\par
    \begin{enumerate}[leftmargin=*]
        
        \item \textit{New features in ISA and lack of stable toolchain:} Earlier studies assume that ISA is independent of hardware implementation, but the data leakage analysis shows that ISA can be exploited as side-channels. The first step is to examine the applicability of current attack vectors in RISC-V processors and then study the new features of RISC-V ISA, which introduces new variables to SCAs. To this aim, a reliable and stable toolchain is required. 
        \item \textit{Need for generic defense mechanisms:} RISC-V, as a new open-source ISA, has a high potential to be the platform for state-of-the-art defense mechanisms. However, our observations show that most of the proposed techniques utilize special features of processors, such as Intel’s x86 and ARM, which are not available in RISC-V processors. It will be worthy of proposing generic methodologies for addressing the gap of security studies between software and hardware by utilizing RISC-V capabilities. These defense mechanisms in the early stages of design can be adopted even for lightweight processors and are not limited to complicated high-performance multi-cores.
    \end{enumerate}

\section*{Acknowledgment}
This work was partially supported by Doctoral College Resilient Embedded Systems which is run jointly by TU Wien's Faculty of Informatics and FH-Technikum Wien.

\end{document}